\documentclass[11pt]{llncs}
\usepackage[latin1]{inputenc}
\usepackage{graphicx}
\usepackage{amssymb}
\usepackage{color}
\usepackage{url}
{$\clubsuit$}
\newcommand{\Neige}{{\sc Sosemanuk}}
\newcommand{\Length}{{10}}
\newcommand{\length}{{ten }}
\newcommand{\Serp}{\emph{Serpent24}}

\begin{document}

\title{\Neige, a fast software-oriented stream cipher\thanks{Work partially supported by the French Ministry of Research RNRT Project ``X-CRYPT'' and by the European Commission via ECRYPT network of excellence IST-2002-507932. }}

\author{C. Berbain \inst{1} \and O. Billet \inst{1} \and A. Canteaut \inst{2} \and N. Courtois \inst{3} \and H. Gilbert \inst{1} \and L. Goubin \inst{4} \and A. Gouget \inst{5} \and L. Granboulan \inst{6} \and C. Lauradoux \inst{2} \and M. Minier \inst{7} \and T. Pornin \inst{8} \and H. Sibert \inst{9}}

\institute{
 Orange Labs, France, \email{ \{come.berbain,olivier.billet,henri.gilbert\}@orange-ftgroup.com}
\and
INRIA-Rocquencourt, projet CODES, France, \email{\{anne.canteaut, cedric.lauradoux\}@inria.fr}
\and
University College of London, UK, \email{n.courtois@ucl.ac.uk}
\and
Université de Versailles, France, \email{
  louis.goubin@prism.uvsq.fr}
\and
Gemalto, France, \email{aline.gouget@gemalto.com}
\and
EADS, France, \email{louis.granboulan@eads.net}
\and
INSA de Lyon, France,
\email{marine.minier@insa-lyon.fr}
\and
Cryptolog International, France, \email{thomas.pornin@cryptolog.com} 
\and
NXP Semiconductors, France, \email{herve.sibert@nxp.com}
}
\date{}

\maketitle

\begin{abstract}
  \Neige\ is a new synchronous software-oriented stream cipher,
  corresponding to Profile~1 of the ECRYPT call for stream cipher
  primitives. Its key length is variable between \(128\) and
  \(256\)~bits. It accommodates a \(128\)-bit initial value. Any key
  length is claimed to achieve \(128\)-bit security.  The \Neige\ 
  cipher uses both some basic design principles from the stream cipher
  SNOW~2.0 and some transformations derived from the block cipher
  SERPENT. \Neige\ aims at improving SNOW~2.0 both from the security
  and from the efficiency points of view. Most notably, it uses a
  faster IV-setup procedure. It also requires a reduced amount of
  static data, yielding better
  performance on several architectures. 
\end{abstract}

\section{Introduction}
This paper presents a proposal for a new synchronous
software-oriented stream cipher, named \Neige. The \Neige\  cipher
uses both basic design principles from the stream cipher
SNOW~2.0~\cite{ej02} and transformations derived from the block
cipher SERPENT~\cite{BAK98}. For this
reason, its name should refer both to SERPENT and SNOW. However, it
is well-known that snow snakes do not exist since snakes either
hibernate or move to warmer climes during the winter.Instead \Neige\  is a
popular sport played by the Eastern Canadian tribes. It consists in
throwing a wooden stick along a snow bank as far as possible. Its
name means snowsnake in the Cree language, since the stick looks like a
snake in the snow. {\em Kwakweco-cime win} is a variant of the same
game but does not sound like an appropriate cipher name. More
details on the \Neige\ game and a demonstration can be found
in~\cite{sose1} and~\cite{sose2}.

The \Neige\ stream cipher is a new synchronous stream cipher dedicated
to software applications. Its key length is variable between \(128\)
and \(256\)~bits. Any key length is claimed to achieve \(128\)-bit
security. It is inspired by the design of SNOW~2.0 which is very
elegant and achieves a very high throughput on a Pentium~4.  \Neige\ 
aims at improving SNOW~2.0 from two respects. First, it avoids some
structural properties which may appear as potential weaknesses, even
if the SNOW~2.0 cipher with a \(128\)-bit key resists all known
attacks. Second, efficiency is improved on several
architectures by reducing the internal state size, thus allowing for a
more direct mapping of data on the processor registers. \Neige\  also
requires a reduced amount of static data; this lower data cache
pressure yields better performance on several architectures. Another
strength of \Neige\ is that its key setup procedure is based on a
reduced version of the well-known block cipher SERPENT, improving
classical initialization procedures both from an efficiency and a
security point of view.

\section{Specification}

\subsection{SERPENT and derivatives}

SERPENT \cite{BAK98} is a block cipher proposed as an AES candidate.
SERPENT operates over blocks of 128~bits which are split into four
32-bit words, which are then combined in so-called ``bitslice''
mode. SERPENT can thus be defined as working over quartets of 32-bit
words. We number SERPENT input and output quartets from 0 to 3, and
write them in the order: $(Y_3, Y_2, Y_1, Y_0)$. $Y_0$ is the least
significant word, and contains the least significant bits of the 32
4-bit inputs to the SERPENT S-boxes. When
SERPENT output is written into 16 bytes, the $Y_i$ values are
written following the little-endian convention (least significant
byte first), and $Y_0$ is output first, then $Y_1$, and so on.

From SERPENT, we define two primitives called \emph{Serpent1} and
\Serp.

\subsubsection{\emph{Serpent1}}

A SERPENT rounds consist of, in that order:
\begin{itemize}
\item a subkey addition, by bitwise exclusive or;
\item S-box application (which is expressed as a set of bitwise
combinations between the four running 32-bit words, in bitslice mode);
\item a linear bijective transformation (which amounts to a few
XORs, shifts and rotations in bitslice mode), see Appendix \ref{serpentt}.
\end{itemize}
\emph{Serpent1} is one round of SERPENT, without the key addition and the linear transformation. SERPENT uses eight distinct S-boxes (see~\ref{sboxserp} for details), numbered from $S_0$ to $S_7$ on
4-bit words. We define \emph{Serpent1} as the application of $S_2$,
in bitslice mode. This is the third S-box layer of SERPENT.
\emph{Serpent1} takes four 32-bit words as input, and provides four
32-bit words as output.

\subsubsection{\Serp}

\Serp\ is SERPENT reduced to \(24\)~rounds, instead of the
\(32\)~rounds of the full version of SERPENT. \Serp\ is equal to the
first \(24\)~rounds of SERPENT, where the last round (the 24th) is
a complete one and includes a complete round with the linear
transformation and an XOR with the 25th subkey. In other words, the
24th round of \Serp\ is thus equivalent to the thirty-second round of
SERPENT, except that it contains the linear transformation and that
the 24th and 25th subkeys are used (32nd and 33rd subkeys in
SERPENT). Thus, the last round equation on Page~224
in~\cite{BAK98} is
\[R_{23}(X) = L\left( {\hat S}_{23}( X \oplus {\hat K}_{23}) \right)
\oplus {\hat K}_{24}~.\]

\Serp\  uses only 25 \(128\)-bit subkeys, which are the first
25 subkeys produced by the SERPENT key schedule. In \Neige,
\Serp\ is used for the initialization step, only in encryption
mode. Decryption is not used.

\subsection{The LFSR}

\subsubsection{Underlying finite field}

Most of the stream cipher internal state is held in a LFSR containing
\(\Length\)~elements of \(\mathbb{F}_{2^{32}}\), the field with
\(2^{32}\)~elements. The
elements of \(\mathbb{F}_{2^{32}}\) are represented exactly as in SNOW~2.0. We
recall this representation here. Let $\mathbb{F}_2$ denote
the finite
field with \(2\)~elements.
Let $\beta$ be a root of the primitive polynomial:
\[
    Q(X)=X^8+X^7+X^5+X^3+1
\]
on $\mathbb{F}_2[X]$. We define the field $\mathbb{F}_{2^8}$ as
the quotient $\mathbb{F}_2[X]/Q(X)$. Each element in
$\mathbb{F}_{2^8}$ is represented
using the basis $(\beta^7,\beta^6,...\beta,1)$. Since the chosen
polynomial is primitive, then $\beta$ is a multiplicative generator of
all invertible elements of $\mathbb{F}_{2^8}$: every non-zero element
in $\mathbb{F}_{2^8}$ is equal to $\beta^k$ for some integer $k$
($0\leq k\leq 254$).
Any element in $\mathbb{F}_{2^8}$ is identified with an 8-bit integer by the
following bijection:
\[\begin{array}{cccc}
\phi : & \mathbb{F}_{2^8} & \rightarrow & \{0, 1, \ldots, 255\}\\
&x=\sum_{i=0}^7 x_i \beta^i & \mapsto &\sum_{i=0}^7 x_i 2^i
\end{array}
\]
where each $x_i$ is either 0 or 1.
For instance, $\beta^{23}$ is
represented by the integer $\phi(\beta^{23}) = \mathtt{0xE1}$ (in
hexadecimal).Therefore, the addition of two
elements in $\mathbb{F}_{2^8}$ corresponds to a bitwise XOR between
the corresponding integer representations. The multiplication by
$\beta$ is a left shift by one bit of the integer representation,
followed by an XOR with a fixed mask if the most significant bit
dropped by the shift equals~\(1\).

Let $\alpha$ be a root of the primitive polynomial
\[
    P(X)=X^4+\beta^{23}X^3+\beta^{245}X^2+\beta^{48}X+\beta^{239}
\]
on $\mathbb{F}_{2^8}[X]$. The field $\mathbb{F}_{2^{32}}$ is then
defined
as the quotient $\mathbb{F}_{2^{8}}[X]/P(X)$, i.e., its elements
are represented with the basis $(\alpha^3,\alpha^2,\alpha,1)$.
Any element in $\mathbb{F}_{2^{32}}$ is identified with a \(32\)-bit
integer by the following bijection:
\[\begin{array}{cccc}
\psi : & \mathbb{F}_{2^{32}} & \rightarrow & \{0, 1, \ldots, 2^{32}-1\}\\
&y=\sum_{i=0}^3 y_i \alpha^i & \mapsto &\sum_{i=0}^3 \phi(y_i) 2^{8i}
\end{array}
\]
Thus, the addition of two elements in $\mathbb{F}_{2^{32}}$
corresponds to a bitwise XOR between their integer representations.
This operation will hereafter be denoted by $\oplus$. \Neige\  also
uses multiplications and divisions of elements in
$\mathbb{F}_{2^{32}}$ by $\alpha$. Multiplication of $z\in
\mathbb{F}_{2^{32}}$ by $\alpha$ corresponds to a left shift by
\(8\)~bits of \(\psi(z)\), followed by an XOR with a \(32\)-bit mask
which depends only on the most significant byte of $\psi(z)$.
Division of $z\in \mathbb{F}_{2^{32}}$ by $\alpha$ is a right shift
by \(8\)~bits of \(\psi(z)\), followed by an XOR with a \(32\)-bit
mask which depends only on the least significant byte of $\psi(z)$.

\subsubsection{Definition of the LFSR}

\begin{figure}
    \begin{center}
        \resizebox{0.7\hsize}{!}{\input{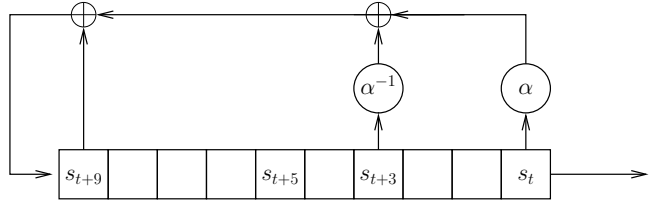}}
    \end{center}
    \caption{\label{fig:lfsr} The LFSR}
\end{figure}

The LFSR operates over elements of $\mathbb{F}_{2^{32}}$. The initial
state, at $t=0$, entails the \length \(32\)-bit values $s_1$ to
$s_{\Length}$. At each step, a new value is computed, with the following
recurrence:
\[
    s_{t+10} = s_{t+9} \oplus \alpha^{-1} s_{t+3} \oplus \alpha s_t,
    \;\; \forall t \geq 1
\]
and the register is shifted (see
Figure~\ref{fig:lfsr} for an illustration of the LFSR).

The LFSR is associated with the following feedback polynomial:
\[
    \pi(X) = \alpha X^{10} + \alpha^{-1} X^7 + X  + 1
        \in \mathbb{F}_{2^{32}}[X]
\]
Since the LFSR is non-singular and since \(\pi\) is a primitive
polynomial, the sequence of \(32\)-bit words \((s_t)_{t \geq 1}\) is
periodic and has maximal period \((2^{320}-1)\).

\subsection{The Finite State Machine}

The Finite State Machine (FSM) is a component with \(64\)~bits of
memory, corresponding to two \(32\)-bit registers $R1$ and $R2$.
At each step, the FSM takes as inputs some words from the LFSR
state; it updates the memory bits and produces a \(32\)-bit output.
The FSM operates on the LFSR state at time~\(t \geq 1\) as follows:
\[FSM_t: (R1_{t-1}, R2_{t-1}, s_{t+1}, s_{t+8}, s_{t+9}) \mapsto (
R1_{t}, R2_{t}, f_{t})\]
where
\begin{eqnarray}
R1_{t} &=& (R2_{t-1} + \mathrm{mux}(\mathrm{lsb}(R1_{t-1}),
        s_{t+1}, s_{t+1}\oplus s_{t+8})) \bmod{2^{32}} \label{eq1}\\
 R2_{t} &=& \mathit{Trans}(R1_{t-1}) \label{eq2}\\
f_{t} & = & (s_{t+9} + R1_{t} \bmod{2^{32}}) \oplus R2_{t}
\label{eq3}
\end{eqnarray}
where $\mathrm{lsb}(x)$ is the least significant bit of $x$,
$\mathrm{mux}(c,x,y)$ is equal to $x$ if $c = 0$, or to $y$ if $c =
1$. The internal transition function \emph{Trans} on
\(\mathbb{F}_{2^{32}}\) is defined by
\[
    \mathit{Trans}(z) = (M \times z \bmod{2^{32}})_{<\!<\!< 7}
\]
where $M$ is the constant value \texttt{0x54655307} (the hexadecimal expression of the first ten decimals of $\pi$) and $<\!<\!<\!$
denotes bitwise rotation of a \(32\)-bit value (by \(7\)~bits here).

\subsection{Output transformation}

\begin{figure}
    \begin{center}
        \resizebox{0.7\hsize}{!}{\input{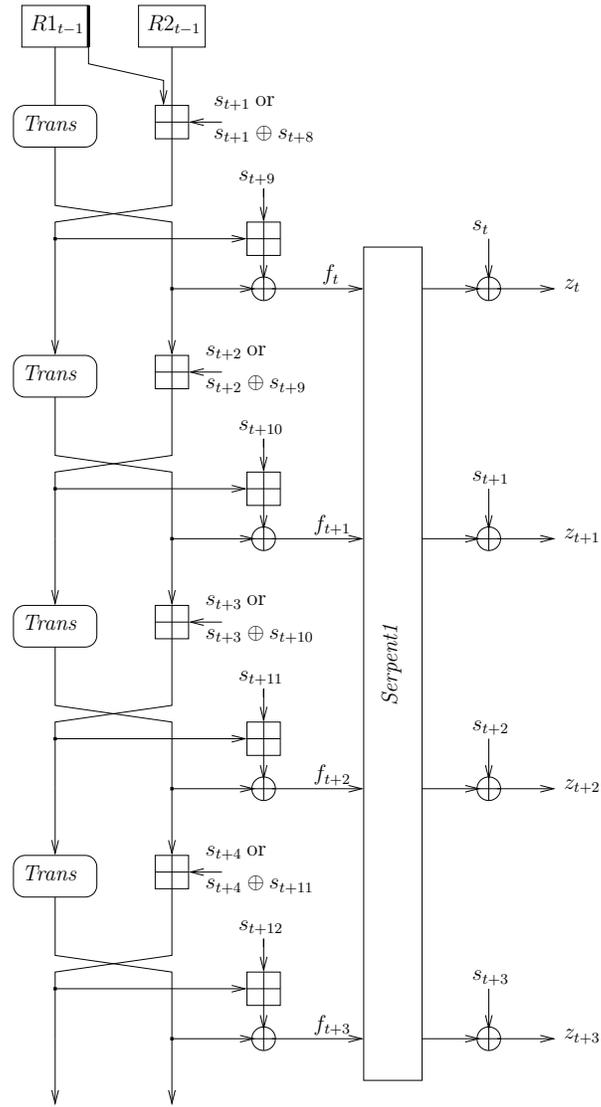}}
    \end{center}
    \caption{\label{fig:feistel} The output transformation on four consecutive rounds of \Neige.}
\end{figure}

The outputs of the FSM are grouped by four, and \emph{Serpent1} is applied to
each group; the result is then combined by XOR with the
corresponding dropped values from the LFSR, to produce the output values
$z_t$:
\[
    (z_{t+3}, z_{t+2}, z_{t+1}, z_t) =
        \mbox{\emph{Serpent1}}(f_{t+3}, f_{t+2}, f_{t+1}, f_t)
        \oplus (s_{t+3}, s_{t+2}, s_{t+1}, s_t)
\]
Four consecutive rounds of \Neige\ are depicted in
Figure~\ref{fig:feistel}.

\subsection{\Neige\  workflow}

The \Neige\  cipher combines the FSM and the LFSR to produce
the output values $z_t$. Time $t=0$ designates the internal state after
initialization; the first output value is $z_1$. Figure~\ref{fig:full}
gives a graphical overview of \Neige.

\begin{figure}
    \begin{center}
        \resizebox{0.7\hsize}{!}{\input{full10.pstex_t}}
    \end{center}
    \caption{\label{fig:full} An overview of \Neige}
\end{figure}

At time $t\geq 1$, we perform the following operations:
\begin{itemize}
\item The FSM is updated: $R1_t$, $R2_t$ and the intermediate value
  $f_t$ are computed from $R1_{t-1}$,
$R2_{t-1}$, $s_{t+1}$, $s_{t+8}$ and $s_{t+9}$.
\item The LFSR is updated: $s_{t+10}$ is computed, from $s_t$, $s_{t+3}$
and $s_{t+9}$. The value $s_t$ is sent to an internal buffer, and the LFSR is
shifted.
\end{itemize}

Once every four steps, four output values $z_t$, $z_{t+1}$, $z_{t+2}$
and $z_{t+3}$ are produced from the accumulated values \(f_t, f_{t+1},
f_{t+2}, f_{t+3}\) and \(s_t, s_{t+1},
s_{t+2}, s_{t+3}\). Thus, \Neige\  produces \(32\)-bit values. We recommend encoding
them into groups of four bytes using the little-endian convention,
because it is faster on the most widely used high-end software platform
(x86-compatible PC), and because SERPENT uses that convention.

Therefore, the first four iterations of \Neige\  are as follows.
\begin{itemize}
\item The LFSR initial state contains values $s_1$ to $s_{\Length}$;
  no value $s_0$ is defined. The FSM initial state contains $R1_0$ and
  $R2_0$.
\item During the
first step, $R1_1$, $R2_1$ and $f_1$ are computed from \(R1_0\),
\(R2_0\), \(s_2\), \(s_9\) and \(s_{10}\).
\item The first step produces the buffered intermediate values $s_1$
and $f_1$.
\item During the first step, the feedback word \(s_{11}\) is computed
  from \(s_{10}\), \(s_4\) and \(s_1\), and the internal state of the
  LFSR is updated, leading to a new state composed of \(s_2\) to \(s_{11}\).
\item The first four output values are $z_1$, $z_2$, $z_3$ and $z_4$,
and are computed using one application of \emph{Serpent1} over
$(f_4, f_3, f_2, f_1)$, whose output is combined by XORs
with $(s_4, s_3, s_2, s_1)$.
\end{itemize}

\subsection{Key initialization and IV injection}

The \Neige\  initialization process is split into two steps:
\begin{itemize}
\item the key schedule, which processes the secret key but does not
depend on the IV; and
\item the IV injection, which uses the output of the key schedule and
the IV. This initializes the stream cipher internal state.
\end{itemize}

\subsubsection{Key schedule}

The key setup corresponds to the \Serp\ key schedule, which
produces 25 128-bit subkeys, as 100 32-bit words. These 25 128-bit
subkeys are identical to the first 25 128-bit subkeys produced by
the plain SERPENT key schedule.

SERPENT accepts any key length from \(1\) to \(256\)~bits; hence,
\Neige\ may work with exactly the same keys. However, since \Neige\  aims
at 128-bit security; its key length must then be at least
\(128\)~bits. Therefore, \(128\) bits is the standard key length.
Any key length from \(128\) bits to \(256\)~bits is supported. But,
the security level still corresponds to \(128\)-bit security. In
other words, using a longer secret key does not guarantee to provide
the security level usually expected from such a key.

\subsubsection{IV injection}

The IV is a \(128\)-bit value.  It is used as input to the \Serp\ block
cipher, as initialized by the key schedule.
\Serp\ consists of 24 rounds and the outputs of the 12th, 18th and 24th rounds are used.
We denote those outputs as follows:
\begin{itemize}
\item $(Y^{12}_3, Y^{12}_2, Y^{12}_1, Y^{12}_0)$: output of the 12th
round;
\item $(Y^{18}_3, Y^{18}_2, Y^{18}_1, Y^{18}_0)$: output of the 18th
round;
\item $(Y^{24}_3, Y^{24}_2, Y^{24}_1, Y^{24}_0)$: output of the 24th
round.
\end{itemize}

The output of each round consists of the four 32-bit words just
after the linear transformation, except for the 24th round, for
which the output is taken just after the addition of the 25th
subkey.

These values are used to initialize the \Neige\  internal state,
with the following values:
\[
    \begin{array}{rcl}
      (s_7, s_8, s_9, s_{10}) &=& (Y^{12}_3, Y^{12}_2, Y^{12}_1, Y^{12}_0)\\
      (s_5, s_6) &=& (Y^{18}_1, Y^{18}_3) \\
      (s_1, s_2, s_3, s_4)&=& (Y^{24}_3, Y^{24}_2, Y^{24}_1, Y^{24}_0) \\
      R1_0 &=& Y^{18}_0 \\
      R2_0 &=& Y^{18}_2

    \end{array}
\]

\section{Design rationale}

\subsection{Key initialization and IV injection}\label{KIIV}

\paragraph{Underlying principle.}
A first property of the initialization process is that it is
split into two distinct steps: the key schedule which does not depend
on the IV, and the IV injection which generates the initial state of
the generator from the IV and from the output of the
key schedule. Then, the IV setup for a fixed key is less
expensive than a complete key setup, improving the common design since
changing the IV is more frequent than changing the secret key.

A second characteristic of \Neige\ is that the IV setup is derived
from the application of a block cipher over the IV. If we consider
the function $F_K$ which maps a $n$-bit IV to the first $n$ bits
of output stream generated from the key $K$ and the IV, then $F_K$
must be computationaly indistinguishable from a random function over
$\mathbb{F}_2^n$. Hence, the computation of $F_K$ cannot ``morally'' be
faster than the best known PRF over $n$-bit blocks. It so happens that
the fastest known PRF use the same implementation techniques that the
fastest known Pseudo-Random Permutations (which are block ciphers), and
amount to the equivalent performance.

Since \Neige\ stream generation is very fast, the generation of $n$
stream bits takes little time compared to a computation of a robust PRP
over a block of $n$ bits. Following this path of reasoning, we decided
to use a block cipher as the fundation of the IV setup for \Neige:
the IV setup itself cannot be much faster than the application of a
block cipher, and the security requirements for that step are much
similar to what is expected from a block cipher.

%A second characteristic of \Neige\  is that the IV setup corresponds
%to a block cipher. Even if this process seems to be quite expensive,
%we claim that, in any stream cipher, the IV setup cannot be much
%faster than a block cipher. Actually, any stream cipher with an
%\(n\)-bit IV can be used for constructing a family of Perfect Random
%Functions  \((F_K)_{K}\) over \({\bf F}_2^n\) as follows: for any
%secret key \(K\), \(F_K\) is the function which associates to the IV
%the first \(n\)~bits of the produced output sequence. Clearly,
%\(F_K\) must be a PRF, otherwise the stream cipher is vulnerable to
%a distinguishing attack. On the other hand, a block cipher
%\(Y=(P_K)_K(X)\) is a PRP for a given key \(K\) and a particular
%plaintext \(X\). We could derive on the first $n$ keystream bits of
%the streamcipher seen as a PRF a particular PRP corresponding to a
%secure blockcipher. So, the generation of the first $n$ bits cannot
%be faster than a block cipher.

\paragraph{Choice of the block cipher.}
The block cipher used in the IV setup is derived from SERPENT for the
following reasons:
\begin{itemize}
\item SERPENT has been thoroughly analyzed during the AES selection
process and its security is well-understood.
\item SERPENT needs no static data tables, and hence adds little or no
data cache pressure.
\item The SERPENT round function is optimized for operation over data
represented as \(32\)-bit words, which is exactly how data is managed
within \Neige. Using SERPENT implies no tedious byte extraction
from 32-bit words, or recombinations into such words.
\item We needed a block cipher for the key schedule and IV injection;
using something other else than AES seems good for ``biodiversity''.
\end{itemize}

\paragraph{Design of \Serp.}
The IV injection uses a reduced version of SERPENT because SERPENT
aimed at \(256\)-bit security, whereas \Neige\  is meant for
\(128\)-bit security. The best linear bias and differential bias for
a \(6\)-round version of SERPENT are $2^{-28}$ and $2^{-58}$
respectively~\cite{BAK98}. Thus, \(12\)~rounds should provide
appropriate security. Twelve more rounds are added in order to
generate enough data (three 128-bit words are needed for
initializing \Neige), hence \(24\) rounds for \Serp. We rely on the
{\Neige\ } core itself to provide some security margins (the output
of \Serp\  is not available directly to the attacker). Two
consecutive outputs of data are spaced with six inner rounds in
order to prevent the existence of relations between the bits of the
initial state and the secret key bits which could be used in an
attack.

\subsection{LFSR}

The SNOW~2.0 LFSR contains 16~elements, which means 512~bits of
internal state. Since we aim only at 128-bit security, we can
accommodate a shorter LFSR. To defeat time-memory-data trade-off
attacks, 256~bits of internal state at least should be used; we wanted
some security margin, hence an LFSR length a bit more than six words.

\paragraph{LFSR length.} \label{boucle}

The LFSR length~\(n\) must be as small as possible: the bigger the state, the more difficult it is to map the state values on the processor registers. Ideally, the total state should fit in the 16 general-purpose registers that the new AMD64 architecture offers.

For efficient LFSR implementation, the LFSR must not be physically
shifted; moving data around contributes nothing to actual security, and
takes time. If $n$ is the LFSR length, then $kn$ steps (for some integer
$k$) must be ``unrolled'', so that at each step only one LFSR cell is
modified. Moreover, since \emph{Serpent1} operates over four successive
output values, $kn$ corresponds to \({\rm lcm}(4, n)\) and it should
be kept as small as possible, since a higher code size increases code
cache pressure.

These considerations led us to $n=8$ or $10$. But, an LFSR of
length eight presents potential weaknesses which may be exploited in a
guess-and-determine attack (see Section~\ref{section:guess}). Therefore, a LFSR of
length~\(10\) is a suitable choice: the 384-bit internal state length
should be enough; only \(20\)~steps need to be unrolled
for an efficient implementation. The total internal state fits in 12
registers, which should map fine on the new AMD64 architecture.

\paragraph{Feedback polynomial.}
The design criteria for the feedback polynomial are similar to those
used in SNOW~2.0. Since the feedback polynomial must be as sparse as possible,
we chose as in SNOW~2.0 a primitive polynomial of the form
\[\pi(X) = c_0 X^{10} + c_a X^{n-a} + c_b X^{n-b} + 1~,\]
where \(0 < a < b< 10\).  The coefficients \(c_0, c_a\) and \(c_b\)
preferably lie in \(\{1, \alpha, \alpha^{-1}\}\) which are the
elements corresponding to an efficient multiplication in \({\mathbb
  F}_{2^{32}}\).  Moreover, \(\{c_0, c_a, c_b\}\) must contain at
least two distinct non-binary elements; otherwise, a multiple of
\(\pi\) with binary coefficients can be easily
constructed~\cite{Ekdakl_Johansson02,HR02}, providing an equation
which holds for each single bit position.

We also want \(a\) and \(b\) to be coprime with the LFSR
length. Otherwise, for instance if \(d = \gcd(a,10) > 1\), the
corresponding recurrence relation
\[s_{t+10} = c_b s_{t+b} + c_a s_{t+a} + c_0 s_t\]
involves three terms of a decimated sequence \((s_{dt+i})_{t > 0}\)
(for some integer~\(i\)), which can be generated by an LFSR of
length~\(n/d\) \cite{Rueppel86}. These conditions led us to \(a=3\)
and \(b=9\).  Since \(a\) and \(b\) are not coprime, \(c_a\) and
\(c_b\) must be different; otherwise, some simplified relations may be
exhibited by manipulating the feedback polynomial as shown
in~\cite{HR02,Can01}. The values \(c_0=\alpha\),
\(c_3=\alpha^{-1}\) and \(c_9=1\) correspond to a suitable primitive
polynomial that fulfills all previously mentioned conditions.

\subsection{FSM}

\paragraph{The \emph{Trans} function.}
The \emph{Trans} function is chosen according to the following
implementation criteria: no static data tables in
order to reduce the cache pressure and the function must be fast on
modern processors. For these reasons, the \emph{Trans} function is
composed of a \(32\)-bit multiplication and a bitwise rotation which
are both very fast.
The \(32\)-bit multiplication provides excellent ``data mixing'' compared to the number
of clock cycles it consumes. The bitwise rotation avoids the existence
of a linear relation between the least-significant bits of the
inputs and the output of the FSM.

The operations involved in the \emph{Trans} functions are
incompatible with the other operations used in the FSM (addition over
$\mathbb{Z}_{2^{32}}$, XOR operation). Actually, mixing operations on
the ring and on the vector space disables associativity and
commutativity laws.  For instance,
\[ \begin{array}{c}
(M \times(R2_{t-1} + s_{t+1}\mathrm{\ mod\ } 2^{32} )
\mathrm{\ mod\ } 2^{32} )_{<\!<\!< 7}\\ \neq \\ (M \times(R2_{t-1})
\mathrm{\ mod\ } 2^{32} )_{<\!<\!< 7} + (M \times( s_{t+1})\mathrm{\
mod\ } 2^{32} )_{<\!<\!< 7} \mathrm{\ mod\ } 2^{32}.
\end{array}
\]

\paragraph{The $\mathrm{mux}$ operation.}
The $\mathrm{mux}$ operation aims at increasing the complexity of fast
correlation and algebraic attacks, since it decimates the FSM input
sequence in an irregular fashion.  Moreover, this operation can be
implemented efficiently with either control bit extension and bitwise operations, or an architecture specific ``conditional
move'' opcode. Modern C compilers know how to perform those
optimizations when compiling the C conditional ternary operator
``\verb+?:+''. This multiplexer is quite fast and requires no jump.

It is fitting that both LFSR elements \(s_{t+c}\) and \(s_{t+d}\)
(with \(c \leq d\)) in the mux operation are not involved in the
recurrence relation. Otherwise the complexity of
guess-and-determine attacksmight be reduced. The distance \((d-c)\) between those
elements must be coprime with the LFSR length since they must not be
expressed as a decimated sequence with a lower linear complexity.
Here, we choose \(d-c = 7\). Finally, it must be impossible for the
inputs of the mux operation at two different steps correspond to the same
element in the LFSR sequence. For this reason, the mux operation
outputs either \(s_{t+c}\) or \(s_{t+c} \oplus s_{t+d}\). If \(s_{t+c}
\oplus s_{t+d}\) is the input of the FSM at time~\(t\), the possible
inputs at time~\((t+d-c)\) are \(s_{t+d}\) and \(s_{t+d} \oplus
s_{t+2d-c}\), which do not match any previous input. It is worth
noticy that this property does not hold anymore if the mux outputs either
\(s_{t+c}\) or \(s_{t+d}\).

\subsection{The output transformation}
The output transformation derived from \textit{Serpent1} aims at
mixing four successive outputs of the FSM in a nonlinear way.
As a consequence, any \(32\)-bit keystream word produced by \Neige\
depends on four consecutive intermediate values \(f_t\). As a result,
recovering any single output of the FSM, \(f_t\), in a
guess-and-determine attack requires the knowledge of at least 
four consecutive words from the LFSR sequence, \(s_{t}, s_{t+1},
s_{t+2}, s_{t+3}\) (see Section~\ref{section:guess} for details).

The following properties have also been taken into account in the
choice of output transformation.
\begin{itemize}
\item Both nonlinear mixing operations involved in \Neige\  (the \textit{Trans}
operation and the \textit{Serpent1} used in bitslice mode) do not
provide any correlation probability or linear property on the least
significant bits that could be used to mount an attack (see Section~\ref{section:corre} for further details).
\item From an algebraic point of view, those operations are combined
to produce nonlinear equations (see Section~\ref{algatt}).
\item No linear relation can be directly exploited on the least
significant bit of the values $(f_t, f_{t+1}, f_{t+2}, f_{t+3})$,
only quadratic equations with more variables than the number of
possible equations (see Section~\ref{section:corre}).

\item The linear relation between $s_t$ and $\mathit{Serpent1}(f_t, f_{t+1},
f_{t+2}, f_{t+3})$ prevents \Neige\  from SQUARE-like attacks.
\end{itemize}

Finally, the fastest SERPENT S-box ($S_2$) has been chosen in \textit{Serpent1}
from an efficiency
 point of view~\cite{Sbox_SERP}. But, $S_2$ also guarantees
 that there is no differential-linear relation
on the least significant bit (the ``most linear'' one in the output of
the FSM).

%%%%%%%%%%%%%%%%%%%%%%

\section{Resistance against known attacks} \label{attack}

Our stream cipher \Neige\  offers a \(128\)-bit security, based on the
following security model.
\subsection{Security model}

The attacker is a probabilistic Turing Machine with access to a black box (oracle) that accepts the following three instructions: \textsc {Reset}, \textsc {Init} with a 128-bit input, \textsc{GetStream} with a 1-bit output.
The attacker's goal is to distinguish with probability $2/3$ between a black box that generates random output, and a black box that implements the stream cipher, where \textsc {Reset} generates a random key, \textsc {Init} initializes the internal state of the stream cipher with a new chosen IV, and \textsc {GetStream} generates the next bit of keystream.
The attacker is allowed to do $2^{128}$ elementary operations, an instruction to the black box being an elementary operation.

This security model falls under remarks made by Hong and Sarkar \cite{Hong:Sarkar:05}, because the precomputation time is not bounded by our model. Therefore our claim is that the 256-bit key variant of \Neige\  provide a 128-bit security.
We do not know of a formal security model that restricts the precomputation time, i.e. that only allows the attacker one of the probabilistic Turing machines that can be built in a reasonable time from the current content of today's computers. Therefore, our claim is that the 128-bit key variant of \Neige, and all variants with larger keys,  provide a 128-bit security against an attacker that is not allowed to benefit from large precomputation.

\medskip

The following sections focus on the security of \Neige\ against known
attacks. It is important to note that the secret key of the cipher cannot be
easily recovered from the initial state of the generator. Once the
initial state is recovered, the attacker is only able to generate the
output sequence for a particular key and a
given IV.
Recovering the secret key or generating the output for a different IV
additionally requires the cost of an attack on \Serp\ with a certain number of
plaintext/ciphertext pairs.

\subsection{Time-memory-data tradeoff attacks}
Due to the choice of the length of the LFSR (more than twice the key
length), the time-memory-data tradeoff attacks described
in~\cite{Bab95,Gol97,BS00} are impracticable. Moreover, since these
TMDTO attacks aim at recovering the internal state of the cipher,
recovering the secret key requires the additional cost of an attack
against \Serp. The
best time-memory data tradeoff attack is the Hellman's one
\cite{Hel80} which aims at recovering a pair \((K,IV)\). For a
\(128\)-bit secret key and a \(128\)-bit IV, its time complexity is
equal to $2^{128}$ cipher operations (see
\cite{Hong:Sarkar:05} for further details).

\subsection{Guess and determine attacks} \label{section:guess}
The main weaknesses of SNOW 1.0 are related to this type of attacks (two at
least have been exhibited \cite{HR02}, \cite{Can01}). They
essentially exploit a particular weakness in the linear recurrence
equation. This does not hold anymore for the new polynomial choice in
SNOW 2.0 and for the polynomial used in \Neige\ which involve non-binary
multiplications by two different constants.
The first attack \cite{HR02} also exploited a ``trick'' coming from the
dependence between the values $R1_{t-1}$ and $R1_t$. This trick is
avoided in SNOW 2.0 (because there is no direct link between
those two register values anymore) and in \Neige.

\bigskip

The best guess and determine attack we have found on \Neige\ is the following.
\begin{itemize}
\item Guess at time $t$, $s_{t}, s_{t+1}, s_{t+2}, s_{t+3}$, $R1_{t-1}$ and $R2_{t-1}$  (6 words).
\item Compute the corresponding outputs of the FSM $(f_t, f_{t+1}, f_{t+2}, f_{t+3})$.
\item Compute $R2_t=Trans(R1_{t-1})$ and $R1_t$ from
  Equation~(\ref{eq1}) if \(\mathrm{lsb}(R1_{t-1})=1\) (this can be
  done only with probability~\(1/2\)).
\item From $f_t=(s_{t+9} + R1_t \bmod{2^{32}}) \oplus R2_t$, compute $s_{t+9}$.
\item Compute $R1_{t+1}$ from the knowledge of both $s_{t+2}$ and
  $s_{t+9}$; compute $R2_{t+1}$. Compute $s_{t+10}$ from \(f_{t+1}\),
  \(R1_{t+1}\) and \(R2_{t+1}\).
\item Compute $R1_{t+2}$ from $s_{t+3}$ and
  $s_{t+10}$; compute $R2_{t+2}$. Compute \(s_{t+11}\) from
  \(f_{t+2}\), \(R1_{t+2}\) and \(R2_{t+2}\). Now, \(s_{t+4}\) can be
  recovered due to the feedback relation at time~\(t+1\):
\[\alpha^{-1} s_{t+4} = s_{t+11} \oplus s_{t+10} \oplus \alpha s_{t+1}~.\]
\item Compute $R1_{t+3}$ from $s_{t+4}$ and
  $s_{t+11}$; compute $R2_{t+2}$. Compute \(s_{t+12}\) from
  \(f_{t+3}\), \(R1_{t+3}\) and \(R2_{t+3}\). Compute \(s_{t+5}\) by
  the feedback relation at time~\(t+2\): 
\[\alpha^{-1} s_{t+5} = s_{t+12} \oplus s_{t+11} \oplus \alpha s_{t+2}~.\]
\end{itemize}

At this point, the LFSR words $s_{t}, s_{t+1}, s_{t+2}, s_{t+3},
s_{t+4}, s_{t+5}, s_{t+9}$ are known. Three elements ($s_{t+6},
s_{t+7}, s_{t+8}$) remain unknown. To complete the full 10 words state
of the LFSR, we need to guess 2 more words, \(s_{t+6}\) and
\(s_{t+7}\) since each \(f_{t+i}\), \(4 \leq i \leq 7\), depends on
all \(4\)~words \(s_{t+4}\), \(s_{t+5}\), \(s_{t+6}\) and
\(s_{t+7}\). Therefore, this attack requires the guess of \(8\)
\(32\)-bit words, leading to a complexity of \(2^{256}\). 

Note that in \cite{AEK05} and in \cite{TS06} the authors respectively proposed two guess and determine attacks against \Neige\ that have a complexity approximatively equal to $2^{226}$ and $2^{224}$ computations. However, as stated in paragraphs 2.6, 3.2 and 4.1, we never intended  to have more than 128-bit security. The internal state of \Neige\ is 384-bit long, which would be bad practice if we aimed at 256-bit  security. Therefore, those guess-and-determine attacks, while being  interesting theoretical studies, do not compromise the security of \Neige.

\subsection{Correlation attacks}\label{section:corre}
In order to find a relevant correlation in \Neige, the following
questions can be addressed:
\begin{itemize}
\item does there exist a linear relation at bit level between
some input and output bits?
\item does there exist a particular
relation between some input bit vector and some output bit vector?
\end{itemize}

In the first case, two linear relations could be exhibited at the bit
level. In the first, the least significant bit of $s_{t+9}$ was  ``conserved'',
since the modular addition over $\mathbb{Z}_{2^{32}}$ is a linear
operation on the least significant bit. The second linear relation
induced by the FSM concerns the least significant bit of $s_{t+1}$ or
of $s_{t+1} \oplus s_{t+8}$ (used to compute $R1_t$) or the seventh
bit of $R2_t$ computed from $s_{t}$ or of $s_{t} \oplus s_{t+7}$.
We here use that $R2_t=\mathit{Trans}(R1_{t-1})$ and
$R1_{t-1}=R2_{t-2} + \left(s_{t} \mbox{ or } (s_{t} \oplus s_{t+7})
\right) \bmod{2^{32}}$.

No linear relation holds after applying $Serpent1$ and there are too many unknown bits to exploit a relation on
the outputs words due to the bitslice design. Moreover, a fast correlation attack seems to be impracticable because the mux operation prevents certainty in the
dependence between the LFSR states and the observed keystream.

\subsection{Distinguishing attacks}
A distinguishing attack by D. Coppersmith, S. Halevi
and C. Jutla (see \cite{BCHJ02}) against the first version of SNOW
used a particular weakness of the feedback polynomial built on a
single multiplication by $\alpha$. This property does not hold for the choice of the new polynomial in SNOW 2.0  and for the
polynomial used in \Neige\  where multiplication by $\alpha^{-1}$ is
also included. 

In \cite{WBC03}, D. Watanabe, A. Biryukov and C. De Cannière
have mounted a new distinguishing attack on SNOW 2.0 with a complexity about $2^{225}$ operations using multiple
linear masking method. They construct 3 different masks $\Gamma_1=\Gamma$, $\Gamma_2=\Gamma
\cdot \alpha$ and $\Gamma_3=\Gamma \cdot \alpha^{-1}$  based on the same
linear relation $\Gamma$.

The linear property deduced from the masks $\Gamma_i$ $(i=1,2$ or $3$) must hold with a high probability on the both following quantities:  $\Gamma_i \cdot S'(x) = \Gamma_i \cdot x$ and $\Gamma_i \cdot z \oplus \Gamma_i \cdot t = \Gamma_i
\cdot( z \boxplus t)$ for $i$=1,2 and 3, where $S'$ is the transition function of the
FSM in SNOW 2.0. In the case of SNOW 2.0, the hardest hypothesis to satisfy is
the first one defined on $y=S'(x)$. In the case of \Neige, we need $Pr(\Gamma_i \cdot
\mathit{Trans}(x) = \Gamma_i \cdot x)_{i=1,2,3}$ to be high. But, we also need that $\forall i=1,2,3$, the relation
$$ (\Gamma'_i,\Gamma'_i,\Gamma'_i,\Gamma'_i) \cdot (x_1, x_2, x_3, x_4) = 
\mathit{Serpent1}((\Gamma_i,\Gamma_i,\Gamma_i,\Gamma_i) \cdot (x_1, x_2, x_3, x_4))~.$$
for some $\Gamma'_i \in \mathbb{F}_2^{32}$, holds with a high probability.

Due to the bitslice design chosen for \emph{Serpent1}, it seems very difficult to find such a mask. Therefore, the attack described in \cite{WBC03} could not be applied directly on \Neige. 

\subsection{Algebraic attacks} \label{algatt}
Let us consider, as in \cite{BG05}, the initial state of the LFSR at bit level:
\[
(s_{10}, \cdots, s_{1}) = (s_{10}^{31}, \cdots, s_{10}^{0},
\cdots,s_{1}^{31}, \cdots, s_{1}^{0})
\]

Then, the outputs of \Neige\  at time $t \geq 1$
could be written:
\[
F^t((s^{10}_{31}, \cdots, s^{1}_{0}))=(z_t, z_{t+1},z_{t+2},
z_{t+3})
\]
where $F$ is a vectorial Boolean function from $\mathbb{F}_2^{320}$
into $\mathbb{F}_2^{128}$ that could be seen as 128 Boolean functions $F_j, \;
\forall j \in [0 .. 127]$ from $\mathbb{F}_2^{320}$ into $\mathbb{F}_2$.

Let us study the degree of an $F_j$ function
depending on a particular bit of the output or on a linear
combination of output bits because it is not possible to directly
compute the algebraic immunity of each function $F_j$ due to the
very large number of variables (320 input bits). We think that the
following remarks prevent the existence of low degree relations between the inputs and the outputs of $F_j$.
\begin{itemize}
\item The output bit $i$ after the modular addition on
$\mathbb{Z}_{2^{32}}$ is of degree $i+1$ (as described in \cite{BS05}).
\item The output bit $i$ after the \emph{Trans} mapping is of degree $i+1-7 \mathrm{\ mod\ } 32, \; \forall i \neq 6$ and equal to 32 for $i=6$ (as described in \cite{BS05}).
\item The $\mathrm{mux}$ operation does not enable to determine with probability one the exact number of bits of the initial state involved in the algebraic relation.
\item The algebraic immunity of the SERPENT S-box $S_2$ at 4-bit word
level is equal to 2 (see \cite{MPC04} for a definition of the algebraic immunity and more details).
\end{itemize}

Under those remarks, we think that an algebraic attack against
\Neige\  is intractable.

\section{Implementation}

The reference C implementation is also an optimized implementation.
When compiled with the \verb+SOSEMANUK_VECTOR+ macro defined, it is
a full program (with its own \verb+main()+ function) which outputs
two detailed test vectors. Since the LFSR length is ten, we unroll the C code on 20 rounds (see \ref{boucle} for details); each test vector contains:
\begin{itemize}
\item A copy of the secret key (a sequence of bytes, expressed in
hexadecimal).
\item The expanded secret key, as described by the SERPENT specification:
the key is expanded to 256 bits, then read as a 256-bit number with the
little endian convention. The test vector outputs that key as a big
hexadecimal number, with some digit grouping.
\item The 25 \Serp\ subkeys, each of them consisting of four
32-bit words (in the $(K_3, K_2, K_1, K_0)$ order).
\item The 128-bit IV, as a sequence of 16 bytes.
\item The IV, once transformed into four 32-bit words, in the
$(I_3, I_2, I_1, I_0)$ order.
\item The initial LFSR state ($s_1$ to $s_{10}$, in that order).
\item The initial FSM state ($R1_0$ and $R2_0$).
\item Ten times the following data:
    \begin{itemize}
    \item Four times the following:
        \begin{itemize}
        \item the new FSM state ($R1_t$ and $R2_t$);
        \item the new LFSR state, after the update (the dropped
        value $s_t$ is also output);
        \item the intermediate output $f_t$.
        \end{itemize}
    \item The \emph{Serpent1} input.
    \item The \emph{Serpent1} output.
    \item 16 bytes of \Neige\  output.
    \end{itemize}
\item The total stream output (160 bytes).
\end{itemize}

\section{Performance}

\subsection{Software implementation}
This section is devoted to the software performance of \Neige. It compares the performance of \Neige\ with the other candidates selected in the Phase 3 (Software Profile), SNOW 2.0 and AES-CTR using the eSTREAM testing framework and the provided reference C implementations \cite{test}. The three tables Table \ref{P4}, Table \ref{AMD} and Table \ref{Alpha} sum up the results (for the keystream generation, the $IV$ setup and the key setup) given in \cite{tes} for three different architectures: an Intel Pentium 4 (CISC target), an AMD Athlon64 X2 4200+ (CISC target) and an Alpha EV6 (RISC target).

 All the results presented for \Neige\  have been computed using the supplied reference C implementation.
 
\paragraph{Code size.}
The main unrolled loop implies a code size between 2 and 5 KB
depending on the platform and the compiler. Therefore, the entire code
fits in the L1 cache. 

\paragraph{Static data.}
The reference C implementation uses static data tables with a total
size equal to 4 KB. This amount is \(3\)~times smaller than the size
of static data required in SNOW~2.0, leading to a lower date cache
pressure. 

\paragraph{Key setup.}
We recall that the key setup (the subkey generation given by
\emph{Serpent24}) is made once and that each new IV injection for a given key corresponds to a small version of the block cipher SERPENT.

The performance of the key setup and of the IV setup in \Neige\ are
directly derived from the performance of SERPENT~\cite{Gladman}. Due
to intellectual property aspects, our
reference implementation does not re-use the best implementation of
SERPENT.
However, the performance given in~\cite{MF05} (i.e., computed on the Gladman's
code written in assembly language \cite{Gladman}) leads to the
following results on a Pentium~4:
\begin{itemize}
\item key setup \(\simeq\) 900 cycles;
\item IV setup \(\simeq\) 480 cycles.
\end{itemize}
These estimations for the IV setup (resp. key setup) performance
corresponds to about \(3/4\)~of the best published performance for
SERPENT encryption (resp. for SERPENT key schedule).

%The key setup in SNOW 2.0 is done for each IV. It is assumed to take around 900~cycles on a Pentium4~\cite{ej02} (the SNOW~2.0 reference implementation provides about 900~cycles on a G4 processor). 

\paragraph{Performance results.}
Table \ref{P4}, Table \ref{AMD} and Table \ref{Alpha} present the performance of the keystream generation (using four performance measures), the agility, the $IV$ setup and the key setup to test the most relevant implementation properties. The four elementary tests for keystream generation are: the encryption rate for long streams by ciphering a long stream in chunks of about 4Kb; the packet encryption rate for three packet lengths (40, 576 and 1500 bytes) including an $IV$ setup; the agility test initiates a large number of sessions (filling 16MB of RAM), and then encrypts streams of plaintexts in short blocks of around 256 bytes, each time jumping from one session to another.

\begin{table}[!ht]
\begin{center}
\begin{tabular}{|c|c|c|c|c|c|c|c|c|c|c|}
\hline
& & & \multicolumn{5}{|c|}{cycles/byte} & cycles/key & cycles/IV \\ \hline
Algo. & Key & $IV$  & Stream & 40 bytes & 576 bytes & 1500 bytes & agility & Key setup& IV setup \\ \hline 
AES CTR & 128 & 128 & 17.81 & 29.19 & 18.35 & 18.04 & 20.77 & 393.45 & 76.16 \\ \hline
SNOW v2.0 & 128 & 128 & 5.04 & 35.60 & 6.92 & 5.92 & 7.95 & 85.44 & 1000.54	\\ \hline
CryptMT (v3) & 128 & 128 & 5.27 & 39.12 & 12.09 & 11.55 & 11.35 & 53.71 & 849.25 \\ \hline
DRAGON & 128 & 128 & 11.37 & 74.09 & 26.07 & 23.23 & 15.00 & 256.04 & 1925.54 \\ \hline
HC-128 & 128 & 128 & 3.76 & 1458.58 & 104.86 & 42.64 & 19.02 & 78.81 & 56929.45 \\ \hline
HC-256 & 128 & 128  & 4.39 & 2596.20 & 184.25 & 73.59 & 26.27 & 76.66 & 104341.33 \\ \hline
LEXv1  & 128 & 128 &  9.46 & 20.78 & 10.88 & 10.01 & 12.30 & 486.57 & 449.00 \\ \hline
NLSv2  & 128 & 128 &  6.64 & 38.94 & 8.52 & 6.97 & 12.10 & 823.74 & 704.68 \\ \hline
Rabbit & 128 & 64 & 9.46 & 34.45 & 11.77 & 10.76 & 12.89 & 984.27 & 825.55 \\ \hline
Salsa20 & 128 & 64 &  16.61 & 42.21 & 17.63 & 18.57 & 18.71 & 90.32 & 78.19 \\ \hline
SOSEMANUK & 128 & 64 & 5.81 & 52.37 & 12.52 & 9.62 & 7.40 & 1287.55 & 1245.71 \\ \hline
\end{tabular}
\end{center}
\caption{\label{P4} Number of CPU cycles for the stream ciphers using a Pentium 4 at 2.80GHz, Model	15/2/9}
\end{table}

\begin{table}[!ht]
\begin{center}
\begin{tabular}{|c|c|c|c|c|c|c|c|c|c|c|}
\hline
& & & \multicolumn{5}{|c|}{cycles/byte} & cycles/key & cycles/IV \\ \hline
Algo. & Key & $IV$  & Stream & 40 bytes & 576 bytes & 1500 bytes & agility & Key setup& IV setup \\ \hline 
AES CTR & 128 & 128 & 13.39 & 18.09 & 13.39 & 13.35 & 15.03 & 152.81 & 15.58 \\ \hline
SNOW v2.0 & 128 & 128 & 4.83 & 23.18 & 5.77 & 5.34 & 6.46 & 43.37 & 528.04	\\ \hline
CryptMT (v3) & 128 & 128 & 4.65 & 19.26 & 8.47 & 7.64 & 8.82 & 25.47 & 384.33 \\ \hline
DRAGON & 128 & 128 & 7.76 & 60.20 & 25.90 & 24.31 & 10.01 & 89.90 & 1449.74 \\ \hline
HC-128 & 128 & 128 & 2.86 & 587.00 & 43.19 & 18.43 & 13.07 & 37.85 & 23308.78 \\ \hline
HC-256 & 128 & 128  & 4.72 & 1420.99 & 103.10 & 42.83 & 21.13 & 41.31 & 56725.89 \\ \hline
LEXv1  & 128 & 128 & 6.84 & 14.19 & 7.78 & 7.20 & 9.19 & 226.41 & 268.31 \\ \hline
NLSv2  & 128 & 128 &  10.69 & 53.24 & 13.45 & 11.48 & 14.13 & 453.35 & 1293.15 \\ \hline
Rabbit & 128 & 64 & 4.98 & 14.60 & 5.55 & 5.25 & 6.34 & 288.21 & 292.38 \\ \hline
Salsa20 & 128 & 64 &  7.64 & 16.10 & 7.74 & 7.91 & 8.93 & 24.57 & 14.29 \\ \hline
SOSEMANUK & 128 & 64 & 4.07 & 25.26 & 7.20 & 6.10 & 5.12 & 759.06 & 560.63 \\ \hline
\end{tabular}
\end{center}
\caption{\label{AMD} Number of CPU cycles for the stream ciphers using an AMD Athlon 64 X2 4200+ at	2.20GHz, Model	15/75/2}
\end{table}

\begin{table}[!ht]
\begin{center}
\begin{tabular}{|c|c|c|c|c|c|c|c|c|c|c|}
\hline
& & & \multicolumn{5}{|c|}{cycles/byte} & cycles/key & cycles/IV \\ \hline
Algo. & Key & $IV$  & Stream & 40 bytes & 576 bytes & 1500 bytes & agility & Key setup& IV setup \\ \hline 
AES CTR & 128 & 128 & 15.53 & 24.63 & 15.94 & 15.82 & 17.80 & 633.65 & 37.58 \\ \hline
SNOW v2.0 & 128 & 128 & 5.17 & 23.74 & 6.11 & 5.73 & 6.37 & 69.00 & 489.35	\\ \hline
CryptMT (v3) & 128 & 128 & 6.90 & 24.74 & 11.64 & 11.75 & 12.86 & 37.49 & 422.17 \\ \hline
DRAGON & 128 & 128 & 8.46 & 74.94 & 41.89 & 40.52 & 10.13 & 234.33 & 1542.46 \\ \hline
HC-128 & 128 & 128 & 3.90 & 1029.93 & 77.41 & 31.59 & 14.80 & 54.67 & 42130.00 \\ \hline
HC-256 & 128 & 128  & 5.18 & 2414.77 & 171.48 & 69.34 & 23.53 & 52.96 & 95937.00 \\ \hline
LEXv1  & 128 & 128 & 	7.99 &	16.87 &	9.15 &	8.44	&	9.53	& 198.49 &	334.58 \\ \hline
NLSv2  & 128 & 128 &  5.93&24.26&6.44&5.59&7.94&530.39&421.66 \\ \hline
Rabbit & 128 & 64 & 5.27 & 14.49 & 5.69 & 5.53 & 6.32 & 318.57 & 280.63\\ \hline
Salsa20 & 128 & 64 & 13.61 & 39.93 & 13.77 & 14.34 & 14.46 & 33.60 & 20.16 \\ \hline
SOSEMANUK & 128 & 64 & 4.63 & 28.80 & 7.66 & 6.26 & 5.32 & 1301.09 & 692.71 \\ \hline
\end{tabular}
\end{center}
\caption{\label{Alpha} Number of CPU cycles for the stream ciphers using an Alpha EV6	at 500MHz, Model	21264}
\end{table}

As shown in these tables, \Neige\ remains among the fastest algorithms on several platforms due
to a good design for the mappings of data on the processor registers
and a low data cache pressure.

\subsection{Hardware implementation}
In \cite{GB07}, the authors propose hardware implementations and performance metrics for several stream cipher candidates and especially \Neige. They remark that even if the design of \Neige\ is a little bit complex to implement, it leads to an impressive performance. The required number of gates for designing \Neige\ on 0.13 $\mu$m Standard Cell CMOS with a key of length 256 bits is 18819 considering that 32 bits are outputted at each cycle. Moreover, the corresponding leakage power is 33.55 $\mu$W for a total power at 10MHz equal to 812.47 $\mu$W. The authors also derive the metrics for maximum clock frequency and for an output rate at 10 Mbps (estimated typical future wireless LAN). In this last case, the corresponding clock frequency is equal to 0.313 MHz for a Power-Area-Time equal to 564.8 nJ-um2. In conclusion, they recommend \Neige\ for WLAN applications with a key length equal to 256 bits. They say that ``with regard to \Neige, the utility as a hardware cipher is clear thus in our opinion requires adding to the hardware focus profile.''

\section{Strengths and advantages of \Neige}
The new synchronous stream cipher \Neige\ based upon the SNOW 2.0
design improves it from several points of view. From a security point
of view, \Neige\ avoids some potential weaknesses as the
distinguishing attack proposed in~\cite{WBC03} due to the particular
use of \emph{Serpent1} in bitslice mode. The chosen LFSR is designed
to eliminate all potential weaknesses (particular decimation
properties, linear relations,...). The mappings used in the Finite
State Machine have been carefully designed in the following way:
\begin{itemize}
\item The \emph{Trans} function guarantees good properties of
  confusion and diffusion for a low cost in software. Moreover, this
  mapping prevents \Neige\ from algebraic attacks.
\item The $\mathrm{mux}$ operation, that could be efficiently
  implemented, protects \Neige\ from fast correlation attacks and
  algebraic attacks.
\end{itemize}

The \emph{Serpent1} output transformation, very efficient in bitslice
mode, provides nonlinear equations, a good diffusion and it improves
the resistance to guess-and-determine attacks. 

The new design chosen for the key setup and the IV injection allows to
split the initialization procedure into two distinct parts, without
any loss of security. It leads to a much faster resynchronization mechanism.

From an efficiency point of view, due to a reduced amount of static
data and a reduced internal state size, the exploitation of the
processor registers is enhanced and the data cache pressure is
improved on several platforms, especially on RISC architectures.

\paragraph{Acknowledgments}
The authors would like to thank Matt Robshaw for valuable comments.

Note that this work was done while the 4th author was affiliated to Axalto/Gemalto (France),
the 7th and the 12th authors were affiliated to France T\'el\'ecom R\&D/Orange
Labs (France), the 8th author was affiliated to the \'Ecole Normale Sup\'erieure
(France), the 10th author was affiliated to INRIA Rocquencourt (France).

%%%%%%%%%%%%%%%%%%%%
%\bibliographystyle{plain}
%\bibliography{stream}

%\newpage
\appendix
%\section{Test vectors}
\section{Specifications of SERPENT}
In this appendix, a recall on the specifications of SERPENT given in \cite{BAK98} is made. First, the S-boxes definition is given and the linear part is also defined again.
\subsection{S-boxes definitions}\label{sboxserp}
The eight SERPENT S-boxes act on 4-bit words and are defined as permutations of $\mathbb{Z}_{16}$:
\[
\begin{array}{lll}
S0 & : & 3, 8, 15, 1, 10, 6, 5, 11, 14, 13, 4, 2, 7, 0, 9, 12 \\
S1 & : & 15, 12, 2, 7, 9, 0, 5, 10, 1, 11, 14, 8, 6, 13, 3, 4 \\
S2 & : & 8, 6, 7, 9, 3, 12, 10, 15, 13, 1, 14, 4, 0, 11, 5, 2 \\
S3 & : & 0, 15, 11, 8, 12, 9, 6, 3, 13, 1, 2, 4, 10, 7, 5, 14 \\ 
S4 & : & 1, 15, 8, 3, 12, 0, 11, 6, 2, 5, 4, 10, 9, 14, 7, 13\\ 
S5 & : & 15, 5, 2, 11, 4, 10, 9, 12, 0, 3, 14, 8, 13, 6, 7, 1\\ 
S6 & : & 7, 2, 12, 5, 8, 4, 6, 11, 14, 9, 1, 15, 13, 3, 10, 0\\
S7 & : & 1, 13, 15, 0, 14, 8, 2, 11, 7, 4, 12, 10, 9, 3, 5, 6
\end{array}
\]
\subsection{Linear part of SERPENT round function}\label{serpentt}
The linear part of a one round version of SERPENT acts on 4 32-bit words $(X_3, X_2, X_1, X_0)$ where $X_0$ is the least significant word and is defined as follows:
\begin{eqnarray*}
X_0 & = & X_0 <\!<\!<\! 13 \\
X_2 & = & X_2 <\!<\!<\! 3 \\
X_1 & = & X_1 \oplus  X_0 \oplus  X_2 \\
X_3 & = & X_3 \oplus  X_2 \oplus  (X_0 <\!<\!<\! 3) \\
X_1 & = & X_1 <\!<\!<\! 1 \\
X_3 & = & X_3 <\!<\!<\! 7 \\
X_0 & = & X_0 \oplus  X_1 \oplus X_3 \\
X_2 & = & X_2 \oplus  X_3 \oplus (X_1 <\!<\!<\! 7) \\
X_0 & = & X_0 <\!<\!<\! 5 \\
X_2 & = & X_2 <\!<\!<\! 22 \\
\end{eqnarray*}

%\end{document}
\end{document}